\documentstyle[fleqn,12pt]{article}

\title{Statistics and Quantum Chaos}
\author{
  Fabio Benatti\thanks{Dept. Theor. Phys.,
                       University of Trieste,
                       Strada Costiera 11, I-34100 Trieste, Italy}, 
  Mark Fannes\thanks{Onderzoeksleider FWO} \\
  Inst. Theor. Phys.,  K.U.Leuven \\
  Celestijnenlaan 200D \\ 
  B-3001 Leuven, Belgium}

\begin{document}
\maketitle

\begin{abstract}
We use multi-time correlation functions of quantum systems
to construct random variables with statistical properties that reflect
the degree of complexity of the underlying quantum dynamics.
\vspace{\baselineskip}
\end{abstract}
\smallskip

\noindent
{\bf PACS numbers:} 05.45.+b, 02.50.-r

\begin{section}{Introduction}
In the field of quantum chaos, the structure of time-correlation
functions is of the utmost importance to study relaxation phenomena, to
single out the existence of different time scales and to perform the
semi-classical  analysis~\cite{CC}. In this letter, we look at time
averages of multi-time correlation functions as expectations of particular
random variables and suggest that their statistical properties might
reflect the degree of irregularity of the quantum dynamics.
In particular, on the level of the fluctuations of these random
variables, a variety of different statistics seems likely to emerge, 
among them the semicircle distribution typical of random matrix theory, 
corresponding to different degrees of randomness. 
\end{section}

\begin{section}{Statistics of correlation functions}
We shall in general be interested in the large-time behaviour of time
correlation functions of quantum dynamical systems. For sake
of comparison and broader generality, such a
matter is better described in the more general setting
of quantum and classical dynamical systems.
 
In quantum mechanics, one usually works with operators $X$ on a Hilbert 
space $\cal H$ and the Heisenberg evolution generated by some
Hamiltonian $H$
\begin{equation}
\label{5}
X\mapsto X(t):= {\rm e}^{i H t}\, X\, {\rm e}^{-i H t}\ . 
\end{equation}
Then, time-invariant expectations 
$X\mapsto \langle X \rangle:=\langle\psi|X |\psi\rangle$
are computed by means of a suitable ``reference state''
$H|\psi\rangle=0$ . In general, that is in the case of discrete quantum
dynamics, $\exp(i H t)$ is replaced by the power $U^n$ of a unitary
operator $U$,  with $U|\psi\rangle=|\psi\rangle$. Then,~(\ref{5})
brings about the time evolution up to an integer multiple $t=Tn$ of a
unit of time $T$.

In classical mechanics, one has a phase-space $\cal X$, a dynamical 
(Hamiltonian) flow connecting phase points $(q,p)$ through trajectories 
$(q_t,p_t)$ and a time-invariant, normalized phase-density distribution 
$\rho$, e.g.\ the canonical ensemble. It is, however, convenient to adopt 
an algebraic description based on the Koopman construction~\cite{RS}, 
using complex-valued functions $f$ on $\cal X$ evolving in time according to
\begin{equation}
\label{7}
f(q,p)\mapsto f_t(q,p):=f(q_t,p_t)\ . 
\end{equation}
Time-invariant expectations are obtained by averaging with respect to $\rho$
\begin{equation}
\label{8}
\langle f\rangle := \int_{\cal X}{\rm d}q{\rm d}p\ \rho(q,p)\ f(q,p)\ .
\end{equation}
The analogy between classical and quantum systems can be pushed  
further by considering the Hilbert space of square integrable functions $f$ 
on phase-space $\cal X$
\begin{equation}
\label{sca}
 \langle f|f\rangle:= \int_{\cal X} {\rm d}q{\rm d}p\ \rho(q,p)\ 
 |f(q,p)|^2\,<\,\infty\ . 
\end{equation}
The classical observables act on this Hilbert space as multiplication operators: 
$f\, |g(q,p)\rangle:= |f(q,p)\, g(q,p)\rangle$ and  we can compute the 
expectation in~(\ref{8}) as the average of such an $f$ with respect to 
the constant ``wave function'' 1 on $\cal X$: 
$\langle f\rangle= \langle 1\,|f|\,1\rangle$. 

In this way, classical and quantum systems can be treated on the 
same footing. Formally, the only difference is the algebra considered: 
commutative in the first case, non-commutative in the second one. 
Nevertheless, this has quite profound consequences on the probabilistic 
structure of the theory. In fact, one of the hard problems for truly 
quantum systems is to understand the implications of positivity, i.e.\ 
$\langle X^\dagger X\rangle\ge 0$, on the structure of the expectations.
Indeed, the algebraic formulation of both classical and quantum dynamical 
systems indicates a possible way to extrapolate from the classical 
to the quantum context, but, at the same time, 
puts into evidence the differences between the two.
For instance, the notion of mixing is expressed for both classical and 
quantum systems by the decorrelation property
\begin{equation}
\label{9}
\lim_t\langle XY(t)Z\rangle = \langle XZ\rangle \langle Y\rangle\ .
\end{equation}
In classical dynamical systems, $Z$ can be commuted over $Y(t)$ so that two 
observables $X$ and $Y$ suffice.
Also, any classical correlation function as $\langle XY(t)ZU(t)VS(t)\rangle$, 
or the like, where the time $t$ appears more than once, can be reduced to the 
form of above.
Because of lack of commutativity this is not possible in quantum system,
unless some form of asymptotic commutativity in time holds as, for instance,
\begin{equation}
\label{10}
  \lim_t \Bigl\langle [X,Y(t)]^* [X,Y(t)]\Bigr\rangle = 0 \ . 
\end{equation}
For infinite quantum systems, many properties can be deduced 
from~(\ref{10})~\cite{T}.

In most finite quantum systems, however, the (quasi-) energy spectrum is 
discrete and neither mixing, nor asymptotic commutativity hold.
Typically, in classically chaotic quantum systems, it is at this point
that the notion of breaking-times appears~\cite{CC}.
We do not want to address this interesting topic here, but we will
stick to quantum systems which are dynamically endowed with some degree of 
asymptotic clustering and show that
different statistics of quantum random variables naturally emerge.
 
In the hierarchy of quantum clustering behaviours, stronger than
mixing is multi-clustering~\cite{NT}
\begin{equation}
  \lim_{\min |t_a- t_b|\to\infty}
  \langle X^{(1)}(t_{i_1})\, X^{(2)}(t_{i_2})\cdots X^{(n)}(t_{i_n})\rangle= 
  \prod_{\ell=1}^s \langle \stackrel{\longrightarrow}{\prod_{j\in J_\ell}} X^{(j)}
  \rangle\ ,
\label{12}
\end{equation}
where $J_\ell$ is the subset $\{j_1,j_2, \ldots\}$ of $\{1,2,\ldots,n\}$ such
that $t_{j_i}= t_{j_k}$, that is we allow a same time to appear
more than once, so that the number of different times $s$ may be smaller than 
the number $n$ of sub-indices.
The short-hand notation $\min|t_a-t_b|\to\infty$ means 
that we let all differences between different times to go to infinity and,
finally, the arrow over the product means that the factors 
inside, which do not in general commute, have to appear in the same order 
as in the correlation function.
It is not difficult to show that~(\ref{12}) is equivalent to having both asymptotic
commutativity~(\ref{10}) and mixing~(\ref{9})~\cite{ABDF}.

However, rather than in a situation where~(\ref{10}) holds, 
we are interested in  multi-time correlation functions of the form
\begin{equation}
\label{13}
 \langle X^{(1)}(t_{i_1})\, X^{(2)}(t_{i_2})\cdots X^{(n)}(t_{i_n})\rangle\ ,
\end{equation}
where $t_{i_1}\ne t_{i_2}\ne\cdots\ne t_{i_n}$,
but, possibly, $t_{i_j}=t_{i_k}$ when $|j-k|>1$.
Indeed, one might rightly suspect that, precisely because of the 
possible irregularity of the dynamics, no asymptotic commutativity is
available
to simplify multi-time correlation functions.
In this case, given several products (monomials) of observables at different times, 
some of them possibly equal, one cannot but keep the monomial as they are,
the only sensible algebraic operation left, apart from linear
combinations and taking adjoints,
being the concatenation of monomials into larger ones.

Concretely, let $X^{(1)},X^{(2)},\ldots,X^{(n)}$ be $n$ 
observables from a suitable operator algebra ${\cal A}$ at time $t=0$. 
Because of the time evolution, in spite of 
possible algebraic relations between $X^{(k)}$ and $X^{(\ell)}$ at $t=0$, 
no simplifying commutation relations, e.g.\ commutativity,
need survive between $X^{(k)}(t_{i_k})$ 
and $X^{(\ell)}(t_{i_\ell})$ for large $|t_{i_k}-t_{i_\ell}|$. 
In general, one will have to cope with expectations of monomials of the form
$X^{(1)}(t_{i_1})X^{(2)}(t_{i_2})\cdots X^{(n)}(t_{i_n})$, the observables at 
time $t=0$ having evolved up to a set of times $t_{i_1},t_{i_2},\ldots,t_{i_n}$, 
some of them possibly equal, without much room for simplifications.
Thus, the only sensible algebraic setting is that provided 
by a ``free product''~\cite{V} of copies of 
the algebra ${\cal A}$  consisting of (linear 
combinations of) monomials $X^{(1)}_{i_1}X^{(2)}_{i_2}\cdots X^{(n)}_{i_n}$, the
sub-index $i_\ell$ locating the observable $X^{(\ell)}$ within the $i_\ell$-th copy of
${\cal A}$, with the following rules

\begin{enumerate}
\item[a)]
 whenever the identity appears it can be dropped;
\item [b)]
 whenever two consecutive observables $X^{(k)}_{i_k}$ and 
 $X^{(k+1)}_{i_{k+1}}$ carry equal subindices ($i_k=i_{k+1}$), then they 
 must be considered as the single observable $\displaystyle 
 \Bigl(X^{(k)}X^{(k+1)}\Bigr)_{i_k}$.
\end{enumerate}

We stress that in the asymptotic free algebra,
monomials are multiplied by concatenation without any simplification rule 
between consecutive letters except for the previous requests a) and b).

We shall call the algebra constructed above the 
``asymptotic free algebra'' and denote it by ${\cal A}_\infty$.
Further, we define an expectation functional $\langle\cdot\rangle_\infty$ on 
the monomials $X^{(1)}_{i_1}\, X^{(2)}_{i_2}\cdots X^{(n)}_{i_n}$ 
by computing consecutive multi-time averages of correlation functions
as in~(\ref{13}), namely
\begin{eqnarray}
\nonumber
 &&\langle X^{(1)}_{i_1}\, X^{(2)}_{i_2}\cdots
 X^{(n)}_{i_n}\rangle_\infty:= \\
 &&\qquad\lim_{T_s} \cdots \lim_{T_1} {1\over T_1 \cdots T_s}
 \sum_{t_s=0}^{T_s} \cdots \sum_{t_1=0}^{T_0}   \langle
 X^{(1)}(t_{i_1})\cdots X^{(n)}(t_{i_n})\rangle 
\label{15}
\end{eqnarray}
in the case of discrete time dynamical systems, otherwise sums have 
to be replaced by integrals.
In the expression above, all time-indices 
$t_{j_\ell}\in\{t_{i_1},t_{i_2},\ldots,t_{i_n}\}$ such that $t_{i_\ell}=t_j$
contibute to the single time average with respect to $t_j$. 
The index $s$ just counts the number of different times that appear
in the multi-time correlation function to be averaged as in~(\ref{15}).
These expectations return positive values  when used to
compute expectations of positive operators and thus the left-hand side
member of~(\ref{15}) allows for a consistent probabilistic
interpretation~\cite{ABDF}. 
Notice that, according to the definition, given an observable $X\in{\cal A}$,
$\langle X_{i_\ell}\rangle_\infty=\langle X \rangle$, whatever the location 
in a $i_\ell$ copy of ${\cal A}$ contributing to  
the asymptotic free algebra ${\cal A}_\infty$.

As a first application, let us assume that multi-clustering~(\ref{12}) holds. 
Then,~(\ref{15}) can be readily
computed yielding 
$$
  \langle X^{(1)}_{i_1}\, X^{(2)}_{i_2}\cdots X^{(n)}_{i_n}
  \rangle_\infty=  \prod_{\ell=1}^k \langle
  \stackrel{\longrightarrow}{\prod_{j\in J_\ell}} X^{(j)} \rangle_\infty 
  = \prod_{\ell=1}^s \langle
  \stackrel{\longrightarrow}{\prod_{j\in J_\ell}} X^{(j)} \rangle\ .
$$
We present explicitly a few expectations choosing for notational simplicity 
$X^{(1)}=A$, $X^{(2)}=B$ and so on.
Remember that the subscripts refer to the times with respect to which the 
limits in~(\ref{15}) are computed, so that equal subscripts 
mean that equal times have been considered:
\begin{eqnarray}
\nonumber
\langle A_1\rangle_\infty&=&\langle A\rangle\\
\nonumber
\langle A_1B_2\rangle_\infty&=&\langle A\rangle\ \langle B\rangle\\
\nonumber
\langle A_1B_2C_1\rangle_\infty&=&\langle AC\rangle\ \langle B\rangle\\
\langle A_1B_2C_1D_2\rangle_\infty&=&\langle AC\rangle\ \langle BD\rangle
\label{17}
\end{eqnarray}
Notice that, because of~(\ref{10}),
we could consistently impose that observables pertaining 
to different copies of ${\cal A}$ in the asymptotic free algebra ${\cal
A}_\infty$ commute, 
that is $[X_k,Y_\ell]=0$ for $k\neq\ell$.
Moreover, the type of clustering into expectations of smaller monomials in~(\ref{17}) 
is an expression of statistical independence of observables well-separated in time.
This has the important consequence that, according to the central limit
theorem, when $N\to\infty$, the fluctuations 
$1/N\sum_{j=1}^N\tilde{A}_j$ in the asymptotic free
algebra ${\cal A}_\infty$ of centered observables $A\in{\cal A}$ 
become Gaussian random variables~\cite{GVV}.

A totally different notion of statistical independence for observables
belonging to the asymptotic free algebra, called ``free independence''
or ``freeness'', is
defined by the following decoupling scheme~\cite{V}
\begin{equation}
\label{18}
\langle X^{(1)}_{i_1}X^{(2)}_{i_2}\cdots X^{(n)}_{i_n}\rangle_\infty=0\ ,
\quad\hbox{if}\quad
\langle X^{(\ell)}\rangle_\infty=\langle X^{(\ell)}\rangle=0\ ,
\end{equation}
with $\ell=1,\ldots,n$ and $i_1\ne i_2\ne \cdots \ne i_n$.  
As a comparison with~(\ref{17}), we write the first few expectations under the
assumtpion that they exhibit free independence.
We let $X^{(1)}=A$, $X^{(2)}=B$ and so on and use~(\ref{18}) after 
writing, say $A$, as $A=\langle A\rangle{\bf 1}+\tilde{A}$ where $\tilde{A}$ 
is now centered.
Then, we notice that, if $A\in{\cal A}$
is centered for $\langle\cdot\rangle$, it is also centered for 
$\langle\cdot\rangle_\infty$ as an observable of the asymptotic free 
algebra $A\in{\cal A}_\infty$.
Thus,
\begin{eqnarray}
\nonumber
\langle A_1\rangle_\infty&=&\langle A\rangle\\
\nonumber
\langle A_1B_2\rangle_\infty&=&\langle A\rangle\ \langle B\rangle\\
\nonumber
\langle A_1B_2C_1\rangle_\infty&=&\langle AC\rangle\ \langle B\rangle\\
\nonumber
\langle A_1B_2C_1D_2\rangle_\infty&=&
\langle AC\rangle\ \langle B\rangle\ \langle D\rangle
+\langle A\rangle\ \langle BD\rangle\ \langle  C\rangle \\
&-&\langle A\rangle\ \langle B\rangle\ \langle C\rangle\ \langle D\rangle
\label{20}
\end{eqnarray}
It follows that the notion of free independence is incompatible with the 
usual statistical independence:
$\langle A_1 B_2 A_1^* B_2^*\rangle_\infty= \langle A A^*\rangle
\langle B B^*\rangle$ in the usual case,  while $\langle A_1 B_2 A_1^*
B_2^*\rangle_\infty= 0$ in the ``free'' case, $A$ and $B$ being centered
observables.  
As a consequence, the fluctuations of centered observables are no longer
Gaussian random variables, but semicircularly distributed~\cite{V}.
\end{section}

\begin{section}{Examples}

We consider a class of dynamical systems described by operators $e(t)$ at
discrete times $t\in {\bf Z}$, $e(t)$ being a unitary operator 
$e= e^\dagger$, $e^2=\bf 1$, specified at time $t=0$ and 
evolved up to time $t$ according to an underlying quantum evolution.
Since we are only interested in the essential features of the time
evolution, like regularity or randomness, we do not take into account
its detailed structure, but rather resort to a schematic description.
We shall assume that the dynamics  may be described by a so-called
``bit-stream''~\cite{Pr,Po,S},  that is by a sequence $a(1),a(2),\ldots$
of  zeroes and ones fixing the commutation relations between operators
at  different times $s,t=1,2,\ldots$
\begin{equation}
\label{2}
 e(s+t)\, e(s) =(-1)^{a(t)}e(s)\, e(s+t)\ .
\end{equation}
Obviously, these commutation relations  strongly depend on the
statistical  properties of the bit-stream.  

The algebra ${\cal A}$ of observables of the system consists of
linear combinations of monomials $w(\bf t)$
of operators $e(t)$ of the form
\begin{equation}
\label{1}
 w({\bf t}):=e(t_{i_1}) e(t_{i_2})\cdots e(t_{i_n})\ ,\qquad
{\bf t}=(t_{i_1}, t_{i_2}, \cdots, t_{i_n})\ .
\end{equation} 
By using the commutation relation~(\ref{2}) and the fact that 
$e(t)^2={\bf 1}$, we may always assume that $\bf t$ is an
ordered multi-index, i.e.\ $t_{i_1}< t_{i_2}< \cdots<t_{i_n}$. 
The probabilities of selfadjoint monomials $w({\bf t})$ are
specified by the expectations $\langle w({\bf t})\rangle$
with respect to a given state $\langle\cdot\rangle$. 

If there are no preferred observables to
single out apart from the identity,  a meaningful statistics arises
from 
\begin{equation}
\label{4}
 \langle w({\bf t})\rangle=0\ ,\quad\langle {\bf 1}\rangle=1\ .
\end{equation} 
The dynamics during a single time step is given by the shift on the indices of the 
operators $e(t)$:
\begin{equation}
\label{3}
 w({\bf t})\mapsto w({\bf t}+1):=e(t_{i_1}+1)\, e(t_{i_2}+1)
 \cdots e(t_{i_n}+1)\ .
\end{equation}
In spite of the extreme simplicity, the variety of statistics brought
about by the expectations in~(\ref{4}) together with the bit-streams
is nevertheless noticeable~\cite{ABDF}.
Notice that $\langle w({\bf t})\rangle$ can appropriately be called a 
multi-time correlation function for the dynamics given in~(\ref{3}).

\begin{subsection}{Free shift}

We shall now consider the so-called ``free shift''. In its most basic
form it is a quantum shift, but without any algebraic
relations as in~(\ref{2}), so that the only possible simplification in
products of observables comes from $e^2={\bf 1}$. It is rather obvious that
system observables do not commute, even when largely separated in time.
The statistics of correlation functions is now described by ``free
independence'', that is by~(\ref{18}).
In order to prove the assertion, we observe that, because of~(\ref{4}),
general centered observables $\tilde A$, 
i.e.\ $\langle \tilde A\rangle=0$, 
are obtained by linear combinations of monomials. 
Since we want to compute expectations of the form  
$\langle w^{(1)}_{i_1}\, w^{(2)}_{i_2}\cdots w^{(n)}_{i_n}\rangle_\infty$,
where $i_j\neq i_{j+1}$ for all $j=\{1,2,\ldots,n\}$, 
we consider time-limits
\begin{equation}
\label{19}
\lim_{\min |t_a- t_b|\to\infty}
\langle w^{(1)}(t_{i_1})\, w^{(2)}(t_{i_2})\cdots w^{(n)}(t_{i_n})\rangle \ ,
\end{equation}
where the $w^{(j)}$ are centered monomials as in~(\ref{1}) 
and $w^{(j)}(t_{i_j})$ are the evolved ones up to times 
$t_{i_j}$ according to~(\ref{3}) and $t_{i_j}\neq t_{i_{j+1}}$.
It is then clear that, for sufficiently large differences between any two consecutive 
times, there cannot be simplifications due to the rule $e(t)^2=\bf 1$. 
Therefore, because of~(\ref{4}), expectations of products of observables 
as in~(\ref{19}) will vanish in the limit, whence
$\langle w^{(1)}_{i_1}\, w^{(2)}_{i_2}\cdots w^{(n)}_{i_n}\rangle_\infty=0$.
\end{subsection}

\begin{subsection}{Regular and irregular quantum shifts}

Quantum shifts governed by generic bit-streams present intermediate situations
interpolating between the case of asymptotic commutativity and the total absence
of algebraic relations between observables largely separated in time.
In fact, one easily calculates 
$$
\langle\Bigl[e(t)\,,e(s)\Bigr]^*\Bigl[e(t)\,,e(s)\Bigr]\rangle =
\Bigl(1-(-1)^{a(|t-s|)}\Bigr)^2\ ,
$$
so that, unless the bit-stream is regular and $\lim_t a(t)=0$ or $1$, there are
no definite commutation relations among operators largely separated in time.
If $a(t)$ is eventually vanishing, then the quantum shift is
asymptotically Abelian and we expect the usual statistical independence for the
(random variables) observables of ${\cal A}_\infty$. If $a(t)$ tends to
1, we obtain Fermionic independence. 
Otherwise, the observables of ${\cal A}$ do not asymptotically
commute and in connection with the degree of irregularity of the
bit-stream, one expects typical statistics to be exhbited in ${\cal A}_\infty$.
Yet, no irregularity is enough to enforce free independence.
In fact, let us consider the simple observable $w=e\in {\cal A}$ and 
monomials constructed with alternating  products of 
$e(t_1)$ and $e(t_2)$, $t_1\ne t_2$. The first, possibly non-zero, 
expectation of such monomials is
$$
  \langle e(t_1) e(t_2) e(t_1) e(t_2) \rangle= (-1)^{a(|t_2-t_1|)} \ . 
$$ 
Since $\langle e_i\rangle_\infty=\langle e\rangle=0$ for all $i$, freeness 
demands $\langle e_1 e_2 e_1 e_2\rangle_\infty=0$.
By choosing a sufficiently irregular bit-stream, e.g.\ a typical path
of an unbiased Bernoulli process, we can enforce
\begin{equation}
 \langle e_1 e_2 e_1 e_2 \rangle_\infty=
 \lim_{T_1,\, T_2\to\infty }
 {1\over T_1 T_2} \sum_{t_1=0}^{T_1} \sum_{t_2=0}^{T_2} 
 (-1)^{a(|t_2-t_1|)}= 0 \ . 
 \label{ce1}
\end{equation}
Nevertheless, using~(\ref{2}) one easily deduces 
$$
  \langle e(t_1)e(t_2)e(t_1)e(t_2)e(t_1)e(t_2)e(t_1)e(t_2)\rangle = 1\ , 
$$
whence $\langle e_1e_2e_1e_2e_1e_2e_1e_2\rangle_\infty = 1$, whereas 
freeness would amount to the vanishing of that expectation, too.

Notice that $\langle (e_1e_2)^4\rangle_\infty$ is the first
expectation of alternating products not to vanish.
By reducing the degree of irregularity of the bit-stream, one may make   
$\langle (e_1e_2)^2\rangle_\infty\neq 0$ in~(\ref{ce1}).
This should be compared with~(\ref{17}) and~(\ref{20}) fixing $C=A$, $D=B$ with
$A$ and $B$ centered observables.
Of course, the full statistics needs the study of higher moments.
However, one may
already guess the connection between the irregularity of the quantum 
dynamics and the clustering of expectations of higher monomials:
the more the randomness the less the contributions.

\end{subsection}

\begin{subsection}{Quantum Koopmanism}

As a somewhat different model, we consider a classical flow $(q,p)\mapsto (q_t,p_t)$ 
with mixing properties on phase-space, namely
$$
 \lim_{t\to\infty} \langle f|g_t\rangle = \lim_{t\to\infty} \int_{\cal X} 
 {\rm d}q{\rm d}p\ \rho(q,p)\ \overline{f(q,p)}\, g(q_t,p_t) 
 = \langle f\rangle\ \langle g\rangle\ ,  
$$
where the Koopman Hilbert space description~(\ref{sca}) for classical systems 
has been used. 
We now proceed to a ``non-canonical'' quantization whereby the quantum
evolution of ``wave-functions'' is exactly the classical one~~\cite{Ber}.
Given functions $f$, $g$ in the Koopman Hilbert space, operators of the 
form $|f\rangle \langle g|$ may be used to construct the non-commutative
algebra of all finite rank operators. 
Expectations of observables $|f\rangle \langle f|$ are given 
by $\langle\, |f\rangle \langle f|\, \rangle= 
|\langle 1\,|f|\,1\rangle|^2$. Finally, the dynamics shifts 
$|f\rangle\langle f|$ into $|f_t\rangle\langle f_t|$ with $f_t$ 
as in~(\ref{7}).
One can then deduce that 
$$
  \lim_{t\to\infty}\langle R(f) R(g_t)\rangle= \langle R(f)\rangle\ 
  \langle R(g)\rangle\ ,
$$
where $R(f):= |f\rangle\langle f|$.  The above product structure
extends to the set of finite rank matrices $A,\, B, \, \ldots, F$ and 
multi-clustering as in~(\ref{12}) holds, namely
$$
  \lim_{\min |t_a- t_b| \to\infty}
  \langle 1|A(t_{i_1})\, B(t_{i_2})\cdots F(t_{i_n})|1\rangle= 
  \langle A\rangle\ \langle B\rangle\ \cdots\ \langle F\rangle\ .
$$
Contrary to~(\ref{12}), there is no clustering of operators
carrying the same time-index.

The above limits can be used to construct the asymptotic 
state $\langle\cdot \rangle_\infty$ on the asymptotic free algebra 
${\cal A}_\infty$. Explicitly 
$$
  \langle A_1B_2\cdots F_n\rangle_\infty=
  \langle A\rangle\ \langle B\rangle\ \cdots\ \langle F\rangle\ .
$$
Notice that the identity operator $\bf 1$ is not a finite rank matrix and, 
in order to construct centered observables
$\tilde{A}:=A-\langle A\rangle\bf 1$, one has to add it 
to the finite rank operators.
Such a dynamical system is neither commutative, nor asymptotically commutative
and therefore, the usual statistical independence~(\ref{12}) and thus a 
Gaussian distribution of fluctuations is
not expected to hold.
Freeness does not show up either as a property of the asymptotic 
free algebra.
Indeed, considering centered observables $\tilde{A}$, $\tilde{B}$,
$\tilde{C}$ and $\tilde{D}$, one can prove that
\begin{eqnarray}
 \nonumber
 \langle \tilde{A}_1\rangle_\infty&=&0\\
 \nonumber
 \langle \tilde{A}_1\tilde{B}_2\rangle_\infty&=&0\\
 \nonumber
 \langle \tilde{A}_1\tilde{B}_2\tilde{C}_1\rangle_\infty&=&\langle A\rangle\ 
 \langle B\rangle\ \langle C\rangle-\langle B\rangle\ \langle AC\rangle\\
 \label{24}
 \langle \tilde{A}_1\tilde{B}_2\tilde{C}_1\tilde{D}_2\rangle_\infty&=&0\ .
\end{eqnarray}
Unlike in~(\ref{17}) and~(\ref{20}) when we use centered observables, 
in~(\ref{24}) the first non-vanishing moment is already the third one, 
which somehow indicates that, despite the mixing property of the underlying 
classical dynamics which is carried over to an exotic quantum dynamics,
the statistics on the asymptotic free algebra does not come nearer to
the irregular quantum shifts discussed above.
Interestingly, the previous way of extending a property of the classical
time evolution, in this case phase space mixing, to a  ``quantum'' system 
was proposed  in~\cite{Ber} to provide a counterexample to the
claimed incompatibility between chaos and quantum mechanics.
Subsequently, a physical application of these ideas was given
in~\cite{Pas}. 
\end{subsection}
\end{section}

\begin{section}{Conclusions}

Usual statistical independence is a workable property in the context of 
infinitely extended dynamical systems
appearing e.g.\ in statistical mechanics where some more or less
strong degree of asymptotic commutativity is expected.
However, when no asymptotic commutativity is available,
the knowledge that multi-time correlation functions, with strictly ordered
times such as in~(\ref{12}), cluster, is not sufficient to draw any conclusion 
about correlation functions where equal times appear as  in~(\ref{13}). 
From the examples of above, we learn that increasing random
behaviours bring us closer to freeness in the sense that more and more
asymptotic expectations vanish. 
This is particularly evident for quantum shifts, where regular bit-streams
would make a lot of multi-time averages return non-zero values.
On the other hand, free independence requires that all expectations of 
monomials of centered observables in the asymptotic free algebra vanish.
This amounts to a total lack of any algebraic structure between observables at 
different times which is difficult to implement by means of whatsoever irregular
bit-stream.
However, freeness seems more likely on the level of the
fluctuations in the asymptotic free algebra of sufficiently random 
quantum systems~\cite{ABDF}.

\end{section}

\noindent
{\bf Acknowledgements:} 
We warmly thank A. Verbeure for many illuminating suggestions
and comments.

One of the authors (F.B.) acknowledges financial
support from the Onderzoeksfonds K.U.Leuven F/97/60 and the
Italian I.N.F.N.

\end{document}